# The Dodona Protocol:
# A Living Design Science Experiment on Blockchain Oracles

Giulio Caldarelli
University of Turin
ORCID ID: 0000-0002-8922-7871
info@dodonaprotocol.org

**V1.12**

**Abstract**

The oracle problem, known as the difficulty of reliably incorporating external information into blockchain-based systems, has been extensively investigated in the academic literature and by practitioners. Recent comparative research has demonstrated that many challenges associated with modern blockchain oracles, including attributability, accountability, integrity, and query design, reflect enduring procedural and epistemic constraints that also characterized ancient oracular institutions such as the Delphic Oracle. The translation of these theoretical insights into applied oracle designs is, however, yet to be explored. This paper introduces the Dodona Protocol, a modular, chain-agnostic oracle service explicitly grounded in the procedural patterns identified through the comparative analysis of ancient and modern oracle systems. Named after the Oracle of Zeus at Dodona, one of the oldest oracular sanctuaries in ancient Greece, the protocol operationalizes several ancient design principles, such as structured consultation calendars, access control, attributable and reputationally accountable resolution, constrained query formats, and donation-weighted service prioritization. The first module implements a query and dispute resolution mechanism in which a named, expert resolver provides binding answers to structured questions submitted by petitioners. The oracle produces outcomes that parties have agreed in advance to accept, rather than claiming access to objective truth. This paper presents the design rationale, architectural overview, and comparative positioning of the Dodona Protocol. The protocol is framed as a living research experiment within the Design Science Research (DSR) tradition, in which the deployed system serves as the research artifact, and operational data inform structured analysis, iterative refinement, and the peer-reviewed dissemination of findings, thereby helping bridge the gap between oracle theory and oracle practice.

## Table of Contents

## I. INTRODUCTION

The oracle problem broadly concerns the difficulty of incorporating external information into blockchain-based systems in a manner that preserves reliability, integrity, and resistance to manipulation [1], [2]. Because oracle mechanisms operate at the interface between on-chain and off-chain environments, they introduce a range of recurring challenges and risks that have been extensively cataloged in the literature [3], [4], [5]. Despite significant technological advances, oracle systems continue to experience attacks and failures, suggesting that the oracle problem may not be completely addressed through technical mechanisms alone [6], [7].

Recent comparative research has offered a complementary perspective. By systematically examining the consultation procedures of the ancient Delphic Oracle alongside modern blockchain oracle designs, Caldarelli and Ornaghi [8] demonstrated that many challenges commonly associated with blockchain oracles such as attributability, accountability, integrity, availability, latency, accessibility, and query design, are not unique to modern technologies but reflect enduring epistemic and procedural constraints inherent to oracle-based systems. The resulting comparative framework identified recurring patterns that transcend technological epochs, suggesting that oracle reliability is fundamentally a procedural and epistemic challenge rather than a purely technological one.

However, the translation of these theoretical findings into applied oracle designs remains largely unexplored in academic as well as practitioner literature. The blockchain oracle landscape is dominated by systems engineered drawing from computer science or economic literature, without systematic reference to the procedural wisdom accumulated across millennia of oracle practice. This paper addresses that gap by proposing designs drawn from ancient oracle practices.

The Dodona Protocol is introduced as a “living research experiment”: a modular, chain-agnostic oracle service whose design choices are explicitly derived from the comparative analysis of ancient and modern oracle systems. Rather than proposing a production-ready platform, this paper establishes the conceptual and architectural foundations of a research-driven oracle that evolves incrementally, module by module, as each component is developed, tested, and refined. This research-driven approach is explicitly grounded in the Design Science Research (DSR) paradigm [9], [10], where the deployed artifact serves as the research instrument: knowledge is generated through cycles of design, deployment, observation, and refinement, with findings disseminated through peer-reviewed channels. The first module of this protocol implements a query and dispute resolution mechanism grounded in the principle of "resolution", as the oracle produces binding outcomes that parties have agreed in advance to accept, without claiming access to objective truth.

The protocol is named after the Oracle of Zeus at Dodona, in Epirus, one of the oldest and most venerated oracular sanctuaries in the ancient Greek world [11]. While the Delphic Oracle provides the procedural framework upon which the protocol draws, the choice of Dodona as a name reflects the project's orientation toward foundational, primal oracle design, returning to first principles rather than iterating on existing blockchain-native architectures. Moreover, derivations of Delphic names are already widespread and often misused in the oracle space, making a distinct designation preferable. The choice is also substantive rather than merely aesthetic: the consultation practice documented at Dodona, where petitioners inscribed their questions on lead tablets, predominantly in binary or alternative form, directly anticipates the constrained query formats adopted by the protocol (see section IV.C).

The remainder of this paper is organized as follows. Section II reviews the relevant background and related work. Section III articulates the design philosophy of the Dodona Protocol. Section IV describes the protocol architecture. Section V provides a comparative analysis against existing oracle and dispute resolution systems. Section VI presents the research methodology underlying the protocol's development. Section VII discusses limitations and open research questions. Section VIII concludes.

## II. BACKGROUND AND RELATED WORK

### *A. The Oracle Problem*

The concept of oracle machines, introduced by Turing [12], describes a computational system with access to an external entity that can instantaneously answer specific questions. Because the machine cannot verify oracle responses, it must treat them as correct, and any dependent computation reflects that blind reliance. In the blockchain context, this dependency has been recognized as a fundamental design challenge since the earliest proposals for smart contracts that interact with external data [13], [14].

Multiple systematizations of knowledge have been proposed for oracle design and related schemes [3], [4], [5], [15]. These contributions have identified core architectural components (data source, communication channel, and contract), recurring risks (centralization, collusion, Sybil attacks, lazy equilibrium, freeloading), and the critical role of query structure in determining oracle reliability [16]. Despite this growing body of research, the oracle problem remains open, and deployed oracle systems continue to exhibit vulnerabilities that suggest a mismatch between theoretical understanding and practical implementation.

### *B. Dispute Resolution and Subjective Oracles*

A distinct strand of oracle research concerns the resolution of disputes and subjective queries, cases where no single authoritative data source exists, and outcomes must be determined through deliberation or voting. Kleros [17] implements a Schelling-point juror system in which token-staked participants independently select outcomes, with economic incentives designed to reward convergence on a focal point. UMA [18] employs an optimistic oracle model in which a proposed answer is accepted unless challenged within a dispute window, with escalation to a broader token-holder vote. Augur [19] combines prediction markets with decentralized reporting to resolve event outcomes.

These systems represent significant contributions to oracle design. However, they share the common assumption that truth or correctness can be approximated through aggregation mechanisms, whether juror consensus, optimistic acceptance, or token-weighted voting. The Dodona Protocol departs from this assumption, as described in Section III.

### *C. The Delphic Comparative Framework*

The theoretical foundation for the Dodona Protocol query response mechanism derives from a comparative framework developed by Caldarelli and Ornaghi [8], which systematically examined the consultation procedures of the ancient Delphic Oracle alongside modern blockchain oracle systems. By reconstructing the Delphic consultation process from historical sources and analyzing 167 historical queries through contemporary oracle taxonomies, the study identified recurring procedural patterns across ancient and modern oracle designs.

Key findings relevant to the present work include the following. First, the Delphic Oracle operated under structured availability constraints (consultations on designated days, during nine months per year), reflecting an implicit trade-off between response complexity and oracle availability that mirrors tiered access models in contemporary oracle systems. Second, oracle accessibility at Delphi required financial commitments (offerings and sacrifices) framed as donations rather than transactional fees, alongside procedural gatekeeping (purification rites) that functioned as eligibility criteria. Third, the Delphic Oracle's reliability was found to increase when queries fell within well-defined epistemic domains, and the lexical analysis confirmed that constrained query structures produced more authoritative and precise responses. Fourth, Delphic integrity mechanisms, including the sealed-urn commit-reveal procedure documented in historical inscriptions, anticipated modern cryptographic commitment schemes. Fifth, the oracle's authority was grounded in a single, attributable source (Apollo, mediated through the Pythia), whose identity was known and whose accountability operated primarily through reputation rather than formal sanctions.

These findings, summarized in Tables 4–6 of [8], provide the design principles that the Dodona Protocol operationalizes, as detailed in the following sections.

## III. DESIGN PHILOSOPHY

### *A. Resolution, Not "Absolute Truth"*

The Dodona Protocol's foundational principle is that the oracle does not claim access to objective truth. Instead, it produces a binding resolution that parties have agreed in advance to accept. This distinction is not merely semantic. It reflects a deliberate epistemic rationale that many real-world queries submitted to oracle systems concern events whose "correct" answer depends on interpretation, framing, or contextual judgment, which Bartholic et al. [16] classify as ambiguous events, where honest and informed parties may legitimately disagree.

The Delphic comparative framework [8] reinforced this insight. No formal accountability mechanism existed at Delphi for incorrect responses. The responsibility for interpretation lay with the petitioner. This accountability structure, not recognized as a deficiency, constituted a design feature as it enabled the oracle to function as a coordination mechanism, providing outcomes that facilitated collective action regardless of their verifiability. The Dodona Protocol adopts an analogous design. Petitioners submit queries with the understanding that the resolver's answer constitutes the final, binding outcome. The oracle's value lies not in omniscience but in the provision of a credible, attributable, and procedurally consistent focal point for resolution. To facilitate straightforward interpretation of the outcome, building on the Delphic sealed-urn design, petitioners are incentivized to provide a priori alternatives for the outcomes to be selected by the oracle instead of requesting an open-ended response.

### *B. Modularity and Incremental Development*

The Dodona Protocol is designed as a modular system. The first module, described in this paper, implements query and dispute resolution, a service in which human expert judgment is the primary mechanism. Future modules, including, DLCs, randomness, automated price feeds and data attestation services, are planned but will be developed and deployed independently. The next planned module extends the protocol to Bitcoin Layer 1 through Discreet Log Contracts (DLCs), enabling Bitcoin-native oracle functionality without requiring intermediary chains or wrapped assets (see Section IV.F). This modularity reflects both pragmatic constraints (a research-driven project develops incrementally) and a theoretical commitment: different classes of oracle queries require fundamentally different resolution mechanisms. Discernible events (publicly observable data such as asset prices) may be resolved through automated aggregation, while ambiguous or sanctioned events (subjective disputes, normative judgments) require human deliberation [8], [16]. This modularity is not limited to the protocol's internal development. The Dodona Protocol is also conceived as a composable base layer on top of which third parties may build independent smart contracts, applications, and complementary modules that depend on Dodona's published outcomes. In this sense, the protocol is intentionally designed not as a closed service, but as an enabling infrastructure for the broader Web3 ecosystem. The development of third-party contracts that condition their execution on Dodona's resolutions is therefore not merely permitted, but actively encouraged, as the protocol exists precisely to support other Web3 systems requiring attributable and procedurally structured oracle outcomes.

### *C. Chain Agnosticism*

The Dodona Protocol is not bound to any specific blockchain. Queries are submitted off-chain through the protocol's website, and answers are published on-chain to one or more public ledgers selected for low transaction costs and data availability. General support donations to the protocol are instead accepted on multiple chains. Unlike query-attached donations, such support donations are not escrowed and are not covered by the refund mechanism, which operates only on the chain hosting the resolution contract. The oracle's value lies in the resolution process, not in the underlying chain infrastructure. Chain agnosticism also reduces dependency on any single network's availability and governance, a concern identified in the broader oracle literature [20]. In the current reference implementation, this principle is operationalized through a dual architecture. The smart contract stores commitments, donation state, and resolution outputs on-chain, while the plaintext query and the related interface logic are handled off-chain through the protocol's web application and backend server. This separation reduces gas costs, avoids unnecessary publication of query text, and preserves a clearer distinction between settlement logic and user interaction.

### *D. Donation-Based Sustainability*

The Dodona Protocol operates on a donation model rather than a fee-for-service model. Petitioners attach a voluntary donation to their query. The donation is retained by the protocol only if an answer is provided and published; otherwise, petitioners may withdraw their funds. This design mirrors the Delphic accessibility model, in which consultation was accompanied by offerings framed as donations to a sacred institution providing a communal service, rather than as transactional fees [8].

The donation model has several implications. It avoids the extractive dynamics of rent-seeking oracle services. It aligns the protocol's incentives with the provision of useful answers (donations are forfeited only upon delivery). And it frames the oracle as a quasi-public good, a research infrastructure sustained by the community it serves, rather than a commercial service maximizing revenue. Moreover, depending on the congestion, even a small donation can lead to an answer, or in cases where congestion is very low, it can be provided with no donations at all. Pending donations may also be increased before the monthly freeze, which provides a simple prioritization signal when demand exceeds capacity. Conversely, the resolver may waive a donation and allow the petitioner to recover it even when an answer is provided, where the circumstances justify such discretion.

The donation model also serves a methodological purpose. A research artifact studied exclusively on testnet would yield limited behavioral data, since petitioners with no value at stake interact with the system differently than those committing real funds. Accordingly, the protocol follows a two-cycle deployment schedule. The first annual cycle operates on testnet, providing baseline calibration data and a controlled environment for protocol refinement; the second cycle, launched in a following cycle, operates on mainnet under the same donation mechanic, enabling a naturalistic comparison between costless and value-committing interactions. The behavioral contrast between testnet and mainnet operation is itself an object of interest within the broader study of oracle reliability and incentive design.

Retained donations cover the protocol's non-trivial operational costs, server infrastructure, smart-contract deployment and auditing, and the dissemination of findings, which would otherwise fall entirely on the principal investigator. Any retained funds are allocated exclusively to these operational and research-related expenses, consistent with the project's non-profit orientation.

The absence of a native token is consistent with this design. Sustainability is pursued through voluntary donations rather than token issuance, speculative dynamics, or token-based incentive schemes.

### *E. The Consultation Calendar*

Queries may be submitted at any time but are resolved on the 18th of each month, during nine active months per year (with three months of inactivity). This design directly operationalizes two Delphic procedural elements identified in [8]. First, the designated consultation day mirrors the Delphic practice of monthly consultations on the seventh day of each month. Second, the seasonal inactivity period mirrors the three winter months during which Apollo was believed to be absent from Delphi, and consultations ceased. The 18th was chosen as a symbolic reference, evoking the structured periodicity of the Delphic calendar while establishing a distinct identity for the protocol. The cooldown period of three months reinforces the rationale of the research-driven protocol and allows parties to improve the design periodically and draw statistics of consultation use and performance. Unlike Delphi, the inactive months fall in summer rather than winter. The underlying principle is preserved, the dormant phase coincides with the season of lowest demand, but the season itself differs: at Delphi this was winter, when Apollo was believed absent, whereas for the Dodona Protocol winter is expected to concentrate the highest oracle demand, making the summer months the natural period of closure.

The consultation calendar serves multiple functions beyond historical allusion. It creates a natural batching mechanism, allowing the resolver to dedicate focused attention to all pending queries simultaneously. It provides a predictable cadence for petitioners to plan their submissions. And it imposes a deliberate latency that functions as a validation window, consistent with the finding from [8] that low latency is desirable for routine oracle queries while higher latency may be preferable for complex or contested requests.

## IV. PROTOCOL ARCHITECTURE

### *A. System Overview*

The Dodona Protocol consists of three principal actors: the petitioner(s), the resolver, and the chain. Petitioners are individuals or parties who submit queries to the oracle. The resolver is a named, identifiable expert who evaluates queries and provides binding answers. The chain is the public ledger on which answers are published, ensuring immutability and verifiability.

The current implementation designates the protocol’s author as the sole responsible for the dispute and query resolution module, although other sources or experts are involved to collect the required data. This is a deliberate design choice, not a scalability constraint. It serves as a specific operationalization of the Delphic attributability principle: the resolver’s identity is known, their domain expertise is verifiable (through their academic publication record), and their accountability operates through reputational mechanisms rather than economic penalties. Future modules (e.g., automated price feeds) may not require human resolution and may employ different trust models. More broadly, the protocol is designed to serve as an oracle layer for external Web3 applications, allowing third parties to build contracts and modules whose execution depends on Dodona's published outcomes.

As a future development, especially under conditions of sustained congestion, the protocol may designate multiple named resolvers in order to increase throughput and improve scalability. Because resolver identities are public and attributable, petitioners may in principle choose whether to address a query to a specific resolver or to submit it to the protocol more generally, allowing assignment according to availability or domain fit.
The current reference implementation has undergone internal security review and testing, including static analysis and adversarial test scenarios; an independent third-party audit has not yet been performed. The contract retains a limited set of administrative functions, documented in the public repository, that are restricted to the resolver address. As with any Web3 application, such review does not eliminate all technical or operational risk. Figure 1 summarizes the architecture of Dodona Protocol module 1.

**Figure 1**. High-level architecture of Module 1 in the Dodona Protocol

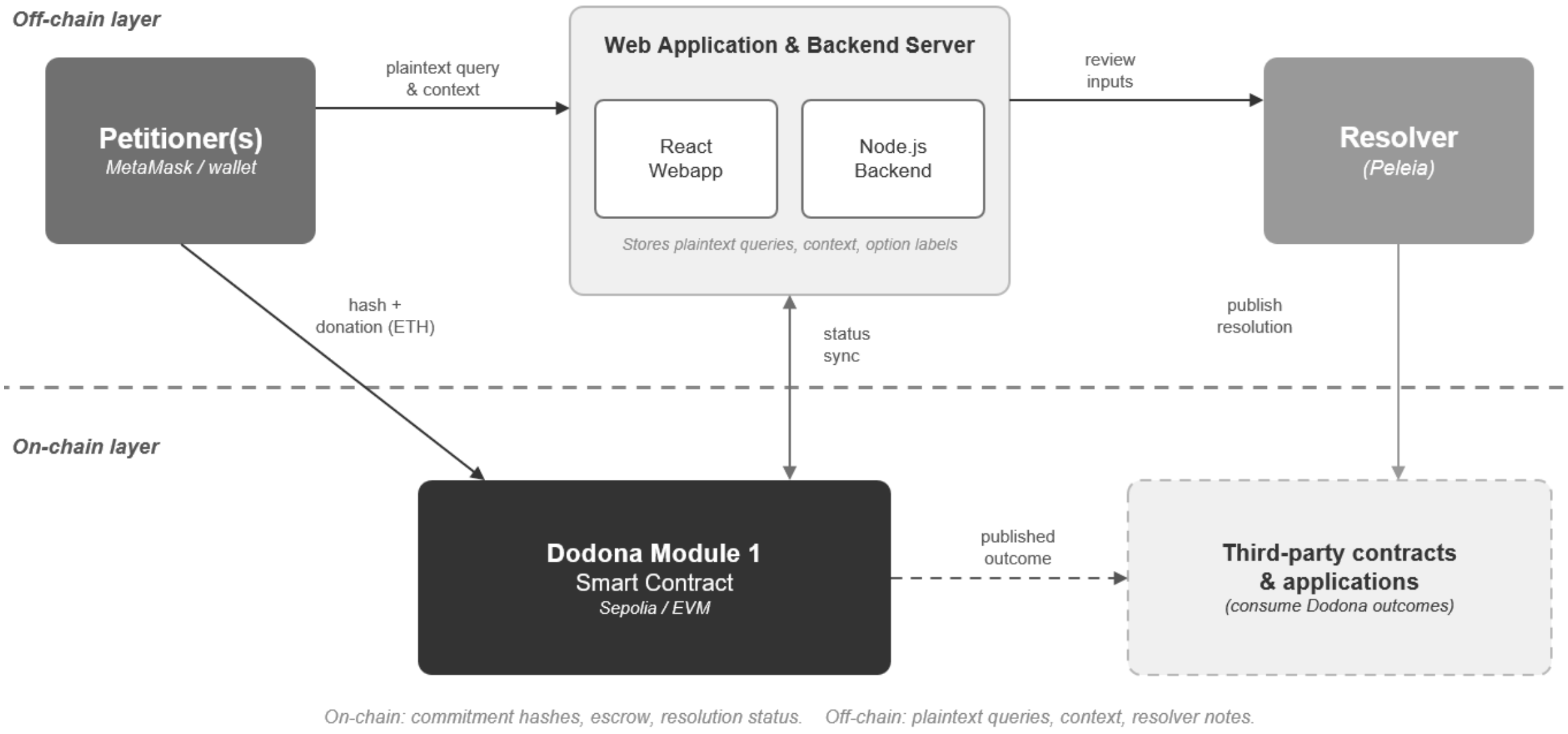

### *B. Query Lifecycle*

The query lifecycle proceeds through the following stages (Figure 2):

**Submission.** Petitioners submit queries through the protocol's off-chain interface (website) at any time during an active month. Each submission includes: (a) the query itself, formulated in a constrained format (see Section IV.C); (b) an attached donation; and (c) any additional context or instructions for the resolver, including the desired answer format (plaintext, hash-encoded, or other encodings specified by the petitioner).

**Escrow.** The attached donation enters a holding state. It is not transferred to the protocol until an answer is published. If no answer is provided by the end of the resolution day (the 18th of the month), the petitioner may withdraw the donation or leave it attached for the next monthly cycle. While the query remains pending, the petitioner may increase the attached donation before the monthly freeze. This also functions as a priority signal in periods of congestion. Priority may also, in future versions of the protocol, reflect predefined partnership arrangements or recognized priority rights attached to specific addresses, particularly where donation levels are equal.

**Review.** On or before the resolution day, the resolver reviews all pending queries. Queries are assessed for answerability, legality, and ethical admissibility (see Section IV.D).

**Freeze**. On the 17th of the month, the pending set for that cycle is frozen on-chain. From that point, withdrawals and donation increases for that monthly round are disabled, and any later submission belongs to the following cycle. The resolver then studies the finalized set off-chain, separating the review phase from the publication phase on the 18th.

**Resolution.** For admissible queries, the resolver provides an answer in the format specified by the petitioner. The answer is published on-chain on the designated public ledger. During this phase, only the resolver interacts with the contract for the frozen monthly batch.

**Settlement.** Upon on-chain publication, the donation is transferred to the protocol. The published answer constitutes the binding resolution. Petitioners who do not receive an answer by the 18th of the month may withdraw their donation or retain it for the next cycle.

**Figure 2**. Queries operational lifecycle

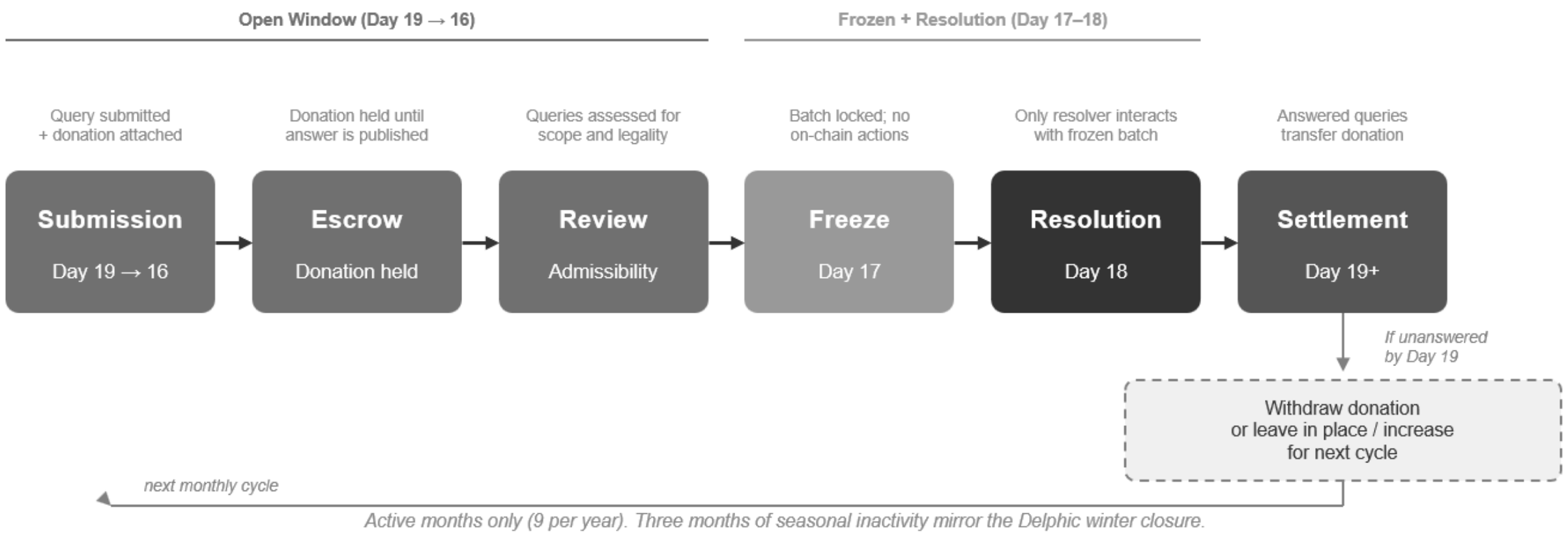

### *C. Query Constraints*

The Dodona Protocol accepts only structured queries with bounded answer spaces. Admissible query formats include: binary queries (yes/no), multiple-choice queries (petitioner-defined option set), bounded numerical ranges, and categorical selections. Open-ended questions are not admissible. For multiple-choice and related formats, the admissible alternatives are committed at submission time, with each option represented by a hash stored on-chain together with the query state.

This constraint is theoretically grounded. The lexical analysis conducted in [8] demonstrated that oracle responses to constrained queries (discernible and sanctioned events) exhibited higher modal density, lower hedge density, and greater precision than responses to open-ended or ambiguous queries. In the blockchain oracle literature, Bartholic et al. [16] similarly observed that oracle performance is particularly sensitive to query structure, with constrained answer spaces reducing interpretive uncertainty and improving reliability. The constraint also has a direct precedent in the sanctuary from which the protocol takes its name The surviving corpus of oracular lead tablets from Dodona records questions posed predominantly in binary or alternative form ("Should I do X?"; "Is it better to do X or to do Y?"), indicating that bounded answer spaces were an established feature of ancient oracular practice long before their formalization in modern oracle taxonomies [21].

Petitioners may also specify the encoding of the answer. For example, a binary query may request that the resolver publish “1” for yes and “2” for no, or may provide specific hash values corresponding to each outcome. This flexibility allows integration with on-chain smart contracts that expect particular data formats, while preserving the resolver’s role as the interpretive authority. In these cases, the resolver does not write free text but resolves by selecting the index of one of the pre-committed options. This structure improves machine readability and makes the module directly usable by third parties, who may write smart contracts conditional on a given oracle outcome (e.g., if the oracle selects option A, contract X executes action Y).

### *D. Refusal Protocol*

The resolver reserves the right to refuse queries that are considered unanswerable (e.g., genuinely open-ended, or requiring information inaccessible to any party) or outside the protocol’s scope. In line with the protocol’s human-mediated nature, ueries are expected to be such that a knowledgeable person could answer them within a bounded amount of time. Questions requiring exhaustive multi-day research, access to proprietary or non-public datasets, or specialized expertise outside the resolver’s competence are considered outside the protocol’s admissibility criteria. This expectation reflects a general principle: the Dodona resolver is a human expert operating within finite time and cognitive resources, not an automated infrastructure or a research delegation service. The resolver also refuses to answer questions that may raise legal or ethical concerns.

Refusals are not silent. When a query is refused, the reason for refusal is published on-chain, and the attached donation may be retained as a penalty. This mechanism serves a dual purpose: it deters frivolous, abusive, or harmful submissions, and it establishes a public, auditable record of the protocol's ethical boundaries. The published refusal reasoning contributes to the protocol's evolving case law, providing precedent for future query assessment. The resolver may additionally blacklist the submitting address from future consultations. While address-based bans are not fully enforceable in pseudonymous environments, the combination of donation forfeiture and public refusal creates a meaningful reputational and financial cost for abusive submissions.

This design mirrors the Delphic gatekeeping function described in [8]. The Pythia could refuse to respond if the consultation prerequisites were not met (e.g., if the sacrificial animal did not exhibit the required signs of divine assent), and historical sources document cases in which petitioners were denied access due to impurity or procedural non-compliance.

### *E. Delphic Procedural Mapping*

Table 1 maps the Dodona Protocol's design elements to the Delphic procedural findings identified in [8], demonstrating the systematic derivation of each architectural choice from the comparative framework.

**TABLE 1. MAPPING OF DODONA PROTOCOL DESIGN TO DELPHIC PROCEDURAL FINDINGS**

| Design Element | Delphic Precedent | Dodona Implementation |
|---|---|---|
| **Attributability** | Apollo as known, trusted source; Pythia as identifiable intermediary | Named resolver with verifiable academic credentials and public identity |
| **Accountability** | Reputational rather than formal sanctions; petitioner bears interpretation risk | Resolver's reputation as primary accountability mechanism; resolution is binding by prior agreement |
| **Availability** | Monthly consultations on designated days; 9 active months per year | Resolution on the 18th of each month; 9 active months, 3 months closed |
| **Accessibility** | Offerings framed as donations; purification as gatekeeping | Donation-based model; refusal protocol as gatekeeping against inadmissible queries |
| **Integrity** | Sealed-urn commit-reveal; oral transmission with procedural safeguards | Petitioner-specified answer encoding (plaintext, hash); on-chain publication for immutability |
| **Query design** | Constrained queries produced more authoritative responses; open-ended queries yielded ambiguity | Structured query formats only (binary, multiple choice, bounded range); open-ended queries refused |
| **Latency** | Low latency is desirable for routine queries; higher latency preferable for complex requests | Monthly batching creates deliberate latency that functions as a review and validation window |

### F. Planned Module: Bitcoin Layer 1 Support via Discreet Log Contracts

The next module planned in the Dodona Protocol roadmap extends oracle functionality to Bitcoin Layer 1 through Discreet Log Contracts (DLCs). DLCs are a class of smart contracts proposed by Dryja [22] that enable two parties to enter into conditional payments on Bitcoin without requiring any modification to the base protocol or any additional trust assumption beyond the oracle itself. In a DLC, the outcome of a contract is determined by an oracle's attestation: the oracle publishes a signed scalar corresponding to the realized outcome, and this signature is used to settle the contract between the parties directly on-chain.

This integration is a natural extension of the Dodona Protocol's design for two reasons. First, the protocol's structured query formats (binary, multiple-choice, bounded range) map directly onto the discrete outcome spaces required by DLCs, making the resolver's index-based resolution mechanism directly compatible with DLC attestation. Second, the named-resolver trust model aligns with DLC architecture, which already assumes a single, identifiable oracle whose public key is known to both counterparties at contract creation. The Dodona resolver's attributable identity and reputational accountability provide a theoretically grounded instantiation of this role.

The Bitcoin L1 module would enable conditional Bitcoin transactions that settle based on Dodona's published resolutions, without requiring the parties to bridge funds to EVM chains or to rely on wrapped assets. This extends the protocol's chain-agnostic principle to the Bitcoin ecosystem and opens the possibility of Bitcoin-native dispute resolution and oracle-dependent contracts settled directly on the Bitcoin base layer. The detailed design, cryptographic specifications, and implementation timeline for this module are subject to ongoing research and remain outside the scope of the present paper.

### G. Planned Module: A Delphic-Inspired Commit-Reveal Oracle

Beyond query and dispute resolution, the Dodona Protocol also contemplates a future module (chronologically planned as the third) concerned with randomness. At a conceptual level, this module is inspired by the Delphic sealed-urn procedure and is included here only to illustrate that the comparative framework may inform oracle designs beyond human-mediated resolution. Rather than being presented as a finalized technical architecture, the randomness module is introduced as a research direction for future development. Its detailed design, implementation choices, and trust assumptions remain outside the scope of the present whitepaper.

## V. COMPARATIVE ANALYSIS

### *A. Positioning in the Oracle Landscape*

The Dodona Protocol occupies a distinct position in the oracle design space. Unlike aggregation-based oracles (Chainlink [23], Pyth, DIA, Band), it does not aggregate data from multiple sources. Unlike optimistic oracles (UMA [18]), it does not rely on challenge windows and token-weighted escalation. Unlike Schelling-point juror systems (Kleros [17]), it does not employ decentralized voting. Instead, it relies on a single, named, expert resolver whose authority derives from attributable identity and verifiable domain expertise.

This positioning is intentional and theoretically grounded. The Delphic comparative framework [8] found that centralization was not inherently problematic when the oracle's identity was known and its reliability could be independently assessed (Table 5 of [8]). The Dodona Protocol operationalizes this finding: the resolver's identity is public, their expertise is documented through academic publications, and their track record of resolutions accumulates over time as a verifiable reputation signal.

### *B. Comparison with Kleros and UMA*

Table 2 compares the Dodona Protocol with Kleros and UMA across key design dimensions.

**TABLE 2. COMPARATIVE ANALYSIS: DODONA VS. KLEROS VS. UMA**

| Dimension | Dodona | Kleros | UMA |
|---|---|---|---|
| **Trust model** | Single expert; reputational accountability | Schelling-point jurors; token-staked voting | Optimistic; challenge window with DVM escalation |
| **Resolver identity** | Named, public | Anonymous jurors | Anonymous proposers; DVM token holders |
| **Latency** | Monthly cycle (deliberate) | Days to weeks (depends on appeals) | Hours (optimistic); days (disputed) |
| **Economic model** | Donation-based; refundable if unanswered | Arbitration fees; juror staking rewards | Bond-based; dispute fees |
| **Query types** | Structured (binary, multiple choice, bounded range) | Open (jurors interpret evidence) | Binary or bounded |
| **Chain dependency** | Agnostic; off-chain submission, on-chain publication | Ethereum-native (with bridges) | Ethereum-native (with L2 support) |
| **Theoretical grounding** | Delphic procedural framework [8] | Schelling focal points; game theory | Optimistic verification; mechanism design |

### *C. What Dodona Deliberately Does Not Do*

The Dodona Protocol makes several deliberate exclusions. It does not claim decentralization; the resolver is a centralized authority, and this is presented as a feature (attributability), not a limitation. It does not provide real-time data feeds; the monthly cadence is unsuitable for applications requiring low-latency price updates. It does not employ token economics, staking, or slashing. More fundamentally, the protocol does not issue, require, or envisage any native token. **No "Dodona Token" exists, nor is one currently planned**. This is a deliberate design choice: a native token could encourage speculative dynamics that are inconsistent with a project conceived exclusively for research purposes. At the current stage, moreover, the implemented modules do not require tokenization to function, as accessibility, prioritization, and sustainability are handled through structured procedures and the donation model rather than through token-based incentives. The protocol does not offer an appeal mechanism; the resolver's answer is final, consistent with the Delphic principle that the oracle's response was authoritative and not subject to renegotiation. However, as in the case of Delphic queries, the petitioner may follow up with another question if not satisfied with the answer.

These exclusions define the protocol's scope and prevent feature creep. The Dodona Protocol is designed to excel at a specific class of oracle queries that are subjective disputes, normative judgments, and ambiguous events that require human expert judgment, rather than competing with existing solutions for automated data feeds.

## VI. RESEARCH METHODOLOGY: A LIVING DESIGN-SCIENCE EXPERIMENT

The Dodona Protocol is presented throughout this paper as a living design science experiment. This section operationalizes that claim by articulating the methodological framework that underpins the protocol's development, the data collection infrastructure that supports empirical analysis, and the iterative cycle through which observations from operation feed back into protocol design and the broader oracle research literature.

### A. Research Paradigm

The Dodona Protocol is developed within the Design Science Research (DSR) tradition, the standard research approach in Information Systems for studying artifacts that are designed, deployed, and observed in real conditions [9], [10]. The premise of DSR is that knowledge can be produced not only by analyzing existing systems but also by deliberately building one and learning from how it behaves over time. Applied to the Dodona Protocol, this premise translates into three commitments: the protocol addresses a concrete and well-documented problem (the open issues in the oracle literature [2], [3], [5], [24]); it is developed and updated incrementally, with each version informed by the previous one; and observations from operation are fed back into the academic discussion, building on the comparative framework introduced in [8].

Within the classification of design-research contributions by Gregor and Hevner [25], the Dodona Protocol falls into the category of work that recombines existing components, commitment schemes, structured query formats, and named arbitration into a new application context, guided by a comparative-historical framework. The contribution is therefore not the invention of new cryptographic mechanisms, but the deliberate translation of historical insights into an operational design.

### B. Data Collection Layer

The off-chain backend server described in Section IV.A is not solely a content delivery layer; it is also designed as a structured data collection infrastructure. Each query interaction generates a record consisting of the submission timestamp, the query category and format (binary, multiple-choice, bounded range), the number and structure of pre-committed options, the donation amount, the resolution outcome (resolved, refused, withdrawn, expired), the time elapsed between submission and resolution, and the resolver’s reasoning where applicable. Refusal records are particularly informative: each refusal is annotated with a category (unanswerable, out of scope, ethically inadmissible, exceeding reasonable resolver capacity) and the textual justification published on-chain.

No directly identifying information is collected. Petitioner addresses are pseudonymous identifiers from the underlying chain; since pseudonymous data are not equivalent to anonymous data under European data-protection law, addresses are processed according to data-minimization principles, stored only in the form already public on-chain, and linked off-chain solely to the query metadata required for the research purposes described above. Participants are informed through the protocol's terms of use that interaction data will be analyzed and published in aggregate form for research purposes, and the submission of a query constitutes acceptance of such use.

### C. The Reflective Cycle: Active and Inactive Months

The protocol’s consultation calendar (nine active months and three inactive months, mirroring the Delphic seasonal pause documented in [8]) acquires a second function under the research framework. The nine active months constitute the operational phase: queries are submitted, resolved, and recorded. The three inactive months are reconceptualized as a reflective phase devoted to systematic analysis of accumulated data, comparison of observed patterns against the existing oracle literature, identification of recurring difficulties or unexpected behaviors, and informed refinement of the protocol for the following annual cycle. This separation of operational and reflective phases is consistent with the iterative structure of DSR [10] and operationalizes the rigor cycle in [9] on a fixed annual cadence.

Table 3 summarizes the annual research cycle, mapping each phase to its activities, methodological tools, and intended outputs.

**TABLE 3. ANNUAL RESEARCH CYCLE OF THE DODONA PROTOCOL**

| Phase | Activities | Methodological Tools | Outputs |
|---|---|---|---|
| **Active phase (9 months)** | Query intake, monthly resolution, on-chain publication, structured logging of submissions, refusals, and resolution metadata | Operational protocol; backend logging; on-chain transaction history; refusal annotations | Resolutions, refusal records, time-series data, evolving on-chain case law |
| **Reflective phase (3 months)** | Quantitative analysis of patterns, qualitative coding of refusals, comparison with literature, design refinement | Descriptive statistics, content analysis, taxonomic coding, comparative-historical analysis | Empirical findings, peer-reviewed publications, protocol-version updates, refined refusal taxonomy |

### D. Analytical Framework

Analysis during the reflective phase follows a mixed-methods approach. The quantitative dimension addresses descriptive questions about query patterns: distribution of query categories (binary, multiple-choice, bounded range), donation distributions, resolution latency, refusal rates by category, withdrawal rates, and recurrence of similar query topics across cycles. These analyses produce a structural characterization of how the protocol is actually used, which can be compared against assumptions in the existing oracle literature [3], [5], [16].

The qualitative dimension focuses on the textual content of the resolver's refusal reasoning and on the categorization of edge cases. Refusal records, published on-chain and accumulated across cycles, are coded into recurring themes (ambiguity that cannot be resolved by structured options, requests requiring inaccessible information, ethically inadmissible queries, queries exceeding reasonable resolver capacity). This coding produces what was earlier described as the protocol's evolving case law: a publicly auditable taxonomy of admissibility boundaries that grows and is refined across cycles. The qualitative track also documents unexpected behaviors and edge cases that suggest design adjustments.

### E. Outputs and Iteration

The reflective phase produces three categories of output. First, empirical findings are written up as peer-reviewed papers and submitted to relevant venues, contributing back to the academic literature on oracle design, dispute resolution, and information-systems research. Second, observed limitations and unexpected behaviors inform versioned refinements of the protocol; smart contracts, query schemas, and refusal categories may be updated for the following annual cycle, with all changes tracked transparently. Third, the accumulated body of refusal reasoning, classification schemes, and analytical findings constitutes a citable corpus that other researchers may draw upon, regardless of whether they intend to use the protocol itself. In this sense, the Dodona Protocol functions not only as an oracle service but also as a research instrument whose outputs are designed to outlive any specific implementation.

This explicit articulation of methodology, data infrastructure, analytical framework, and output expectations is what makes the “living experiment” framing more than a rhetorical device. The protocol generates data, the data feed structured analysis, the analysis informs the next iteration of design, and the findings are released into the academic discussion. The iterative loop is closed by design. Figure 3 provides a description of this research loop.

**Figure 3**. Annual Research Cycle of the Dodona Protocol

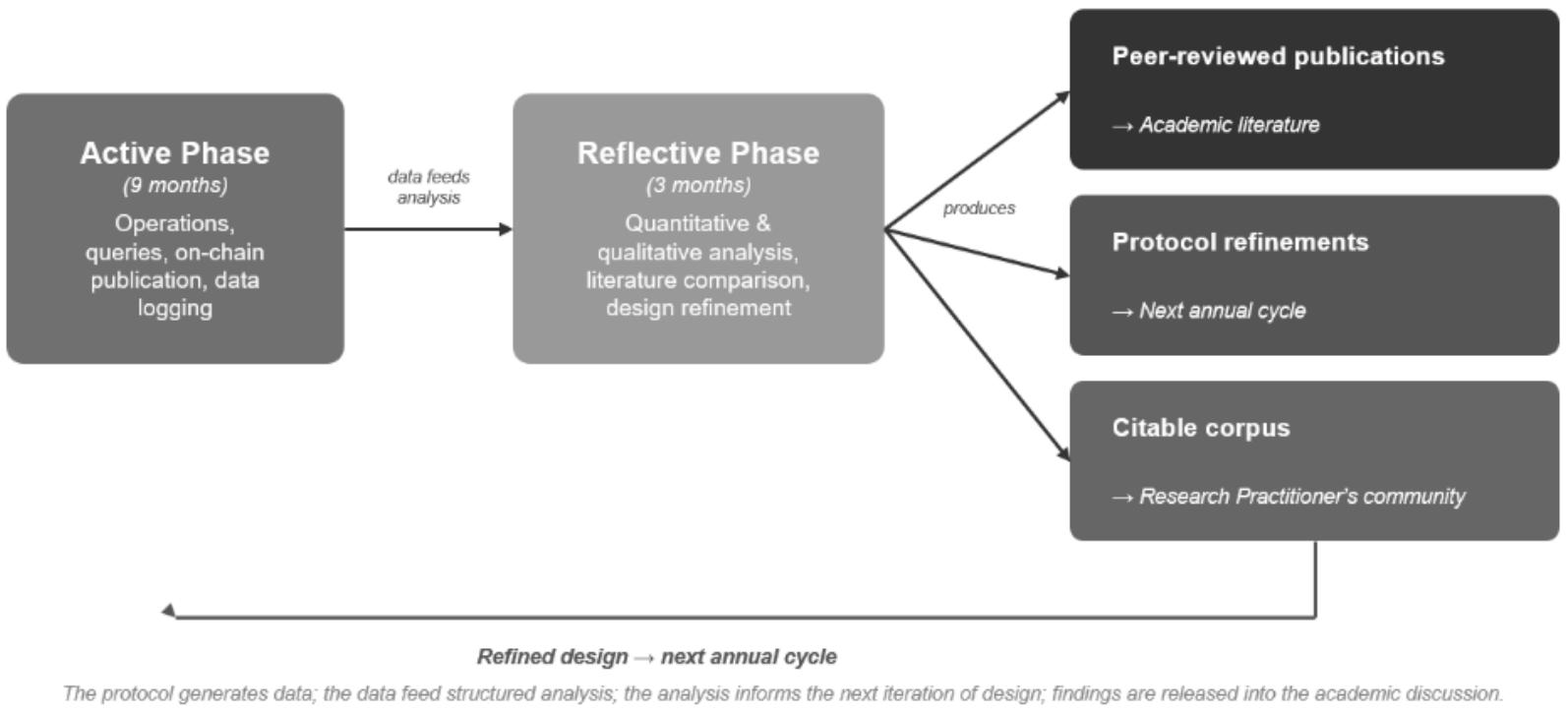

## VII. LIMITATIONS AND OPEN QUESTIONS

The Dodona Protocol’s design entails several acknowledged limitations and open research questions.

**Single point of failure.** The most obvious limitation is the reliance on a single resolver. If the resolver is unavailable, compromised, or acts in bad faith, the protocol has no internal mechanism for correction or appeal. This risk is partially mitigated by the donation refund mechanism (petitioners recover their funds if no answer is provided) and by the public, on-chain nature of all resolutions (which creates an auditable record). However, the protocol does not claim to solve the single-point-of-failure problem; rather, it accepts it as a design trade-off in exchange for attributability and reputational accountability.

**Scalability.** The monthly resolution cycle and single-resolver architecture impose natural throughput constraints. If query volume exceeds the resolver’s capacity, prioritization criteria will need to be developed. The Delphic framework offers precedent here: at Delphi, prioritization was managed through promanteia (priority consultation rights) and, when necessary, drawing lots [8]. Analogous mechanisms may be appropriate for the Dodona Protocol as demand grows. As a base criterion, priority is handled by donation weight (higher donation=higher priority). A further avenue for addressing congestion is the introduction of multiple publicly identified resolvers, echoing the expansion of human mediation that characterized ancient oracular institutions in periods of higher demand. In such a configuration, petitioners could either submit a query to a specific resolver or delegate the choice to the protocol, depending on the desired expertise, availability, or trust preference. In addition to donation-based prioritization, the protocol may in the future recognize priority rights for selected partner addresses or categories of participants. This would function analogously to the Delphic practice of promanteia, under which specific individuals or communities could obtain precedence in consultation order. In the Dodona context, such priority would not necessarily override larger donations, but could operate as a tie-breaking or preference mechanism among queries carrying equivalent donation levels. It is important to note that there is an evident limit to the number of questions a resolver can handle in one session. In line with the oracle at Delphi, the protocol may reasonably handle around 15–20 queries per session (per resolver).

**Adversarial assumptions.** The protocol assumes that petitioners submit queries in good faith and that the resolver acts with integrity. Adversarial scenarios, including collusion between petitioners, resolver corruption, and deliberate submission of trap questions designed to damage the resolver's reputation, remain underexplored and represent important directions for future work. As a standard approach, the responder refuses to answer any question that is sought to harm its or the protocol's reputation.

**Future modules.** The architecture for future modules (randomness, automated price feeds, data attestation) is not specified in this paper. The next planned module, Bitcoin Layer 1 support via Discreet Log Contracts (Section IV.F), is outlined at a conceptual level, but its cryptographic and implementation details remain to be fully specified. The transition from human-resolved to automated modules will introduce fundamentally different trust models, incentive structures, and latency requirements. The modular design is intended to accommodate this heterogeneity, but the integration interfaces between modules remain to be defined. While the randomness module is introduced in Section IV.G, drawing on the Delphic sealed-urn procedure formalized in [8], its on-chain implementation remains to be developed and tested.

**Governance.** The protocol currently has no formal governance structure beyond the resolver's discretion. As the protocol evolves, questions of parameter governance (e.g., which chains to publish on, which months to designate as inactive, donation thresholds) will need to be addressed. Whether governance should remain centralized (consistent with the single-resolver model) or evolve toward more participatory structures is an open question.

**Legal and regulatory considerations.** The provision of binding dispute resolution raises potential legal implications that vary across jurisdictions. The Dodona Protocol does not constitute a legal arbitration service, and its resolutions have no legal force beyond the voluntary agreement of the parties. Nonetheless, the intersection of on-chain oracle services and dispute resolution law represents an emerging area of inquiry that warrants careful attention. The base rationale is that the protocol has to be interpreted as a research experiment only, devoted to the improvement of oracle designs, and its function is not to be used to enforce any legal claim or action. Users interact with it entirely at their own risk. Service availability may be interrupted, suspended, or degraded without prior notice, whether because of smart-contract issues, wallet-provider behavior, backend outages, network congestion, or other forms of technical disruption. Accordingly, no guarantee is made as to uninterrupted operation, and no responsibility is assumed for losses, delays, or disruptions that may in any way affect user interactions with the protocol or associated funds.

## VIII. CONCLUSION

This paper has introduced the Dodona Protocol, a modular, chain-agnostic oracle service explicitly grounded in the procedural patterns identified through the comparative analysis of ancient and modern oracle systems. By systematically mapping the findings of the Delphic comparative framework [8] onto applied oracle design, the protocol demonstrates that theoretical insights from the oracle literature, including structured consultation calendars, donation-based accessibility, attributable resolution, constrained query formats, and donation-weighted service prioritization, can be operationalized in a coherent architectural proposal.

The protocol's first module implements query and dispute resolution through a named, expert resolver who provides binding answers to structured questions. The guiding philosophy of resolution, not truth, reflects an epistemic stance that prioritizes procedural consistency and coordination utility over claims to objective correctness. This orientation positions the Dodona Protocol not as a competitor to existing aggregation-based or voting-based oracle systems, but as a complementary approach suited to the class of queries that require human expert judgment. A randomness module inspired by the Delphic sealed-urn procedure is identified as a direction for future development (Section IV.G), suggesting that the comparative framework is generative and may inform oracle designs beyond human-mediated resolution.

The Dodona Protocol is presented as a living research experiment, evolving incrementally as each module is developed, tested, and refined. This trajectory is structured within the Design Science Research paradigm [9], [10], which provides the methodological grounding for studying the protocol as an evolving research artifact rather than as a finished product. Its contribution lies not in claiming a finished product, but in showing that the gap between oracle theory and practice can be bridged, and that the procedural wisdom accumulated across more than two millennia of oracle systems still offers meaningful insights for contemporary blockchain oracle design. It is not intended to compete with existing oracle services or to enter the market as a commercial product. If a donation-based model is implemented, its sole function would be to support research, development, and the dissemination of results, rather than to generate profit. In this sense, the protocol is offered to the research community as a citable reference for its design principles, a testbed for oracle mechanism design, and an open invitation to reflect on what the oldest human institutions for managing uncertainty can still teach the newest ones. At the time of writing, a reference implementation of Module 1, comprising a Solidity smart contract, a React web application, and a Node.js backend, has been deployed on the Ethereum Sepolia testnet and is publicly accessible for inspection and testing.

**Code and Implementation Availability.** The Dodona Protocol's reference implementation, including the deployed smart contract, the web application source code, and supporting documentation, is publicly accessible at https://dodonaprotocol.org. The Module 1 contract is deployed on the Ethereum Sepolia testnet at address:

0x43259D9A08b13aF95d2dcdCfc9a241CBD2855bB1

the contract ABI, the event signatures emitted upon resolution, and minimal integration examples for third-party contracts are available in the project repository, so that external developers can condition contract execution on Dodona's published outcomes without reverse-engineering the deployment. Readers and researchers are invited to inspect, test, and engage with the artifact as part of the protocol's open, ongoing research process.

**DISCLAIMER**

The Dodona Protocol is developed by the author in a personal and independent capacity, within the Design Science Research paradigm, where the deployed system serves as a research artifact for the collection of empirical data and the production of academic publications. It does not constitute an initiative, product, service, or official activity of the University of Turin, and any academic affiliation reported in this paper serves solely to identify the author's scholarly background and does not imply institutional endorsement, supervision, or responsibility.